\documentclass[12pt]{article}
\usepackage{amsmath,amssymb,amsthm,amsxtra,overpic,bbm,bm,epsfig,subfigure}
\usepackage{hyperref}
\usepackage{mathrsfs}
\usepackage{enumitem}
\usepackage{graphicx}
\usepackage{color}
\usepackage{comment}
\usepackage{epstopdf}
\usepackage{float}
\usepackage{cite}
\usepackage{slashed,stmaryrd}

\def\thefootnote{\fnsymbol{footnote}}

\begin{document}

\vspace{0.2cm}

\begin{center}
{\Large\bf The number of sufficient and necessary conditions for CP conservation with Majorana neutrinos: three or four?}
\end{center}

\vspace{0.2cm}

\begin{center}
{\bf Bingrong Yu}~\footnote{E-mail: yubr@ihep.ac.cn},
\quad
{\bf Shun Zhou}~\footnote{E-mail: zhoush@ihep.ac.cn}
\\
\vspace{0.2cm}
{\small
Institute of High Energy Physics, Chinese Academy of Sciences, Beijing 100049, China\\
School of Physical Sciences, University of Chinese Academy of Sciences, Beijing 100049, China\\}
\end{center}

\vspace{1.5cm}

\begin{abstract}
As is well-known, there exist totally three CP-violating phases in the leptonic sector if three ordinary neutrinos are massive Majorana particles. In this short note, we raise the question whether the number of sufficient and necessary conditions for CP conservation in the leptonic sector with massive Majorana neutrinos is three or four. An intuitive answer to this question would be three, which is also the total number of independent CP-violating phases. However, we give a counter example, in which three conditions are in general not sufficient for CP conservation. Only for all the lepton masses and mixing angles within their experimentally allowed ranges can we demonstrate that it is possible to find out three weak-basis invariants, which should be vanishing to guarantee leptonic CP conservation.
\end{abstract}

%\begin{flushleft}
%\hspace{0.8cm} PACS number(s):
%\end{flushleft}

\def\thefootnote{\arabic{footnote}}
\setcounter{footnote}{0}
\newpage
%\tableofcontents
\newpage
\section{Introduction}\label{sec:intro}
At the low-energy scale, lepton masses, flavor mixing angles and CP-violating phases are governed by the following effective Lagrangian~\cite{Xing:2011zza}
\begin{eqnarray}
-{\cal L}^{}_{\rm mass} = \overline{l^{}_{\rm L}} M^{}_l l^{}_{\rm R} + \frac{1}{2} \overline{\nu^{}_{\rm L}} M^{}_\nu \nu^{\rm C}_{\rm L} + {\rm h.c.} \; , \label{eq:lag}
%     (1)
\end{eqnarray}
where $\nu^{\rm C}_{\rm L} \equiv {\cal C} \overline{\nu^{}_{\rm L}}^{\rm T}$ with ${\cal C} \equiv {\rm i}\gamma^2 \gamma^0$ being the charge-conjugation matrix, $M^{}_l$ and $M^{}_\nu$ stand for the charged-lepton mass matrix and the Majorana neutrino mass matrix, respectively. Long time ago, it was suggested in Ref.~\cite{Dreiner:2007yz} that three vanishing weak-basis (WB) invariants
\begin{eqnarray}
{\cal I}^{}_1 &\equiv& {\rm Tr}\left\{ \left[H^{}_\nu, H^{}_l \right]^3\right\} = 0 \; ,
 \label{eq:I1} \\
%     (2)
{\cal I}^{}_2 &\equiv& {\rm Im}\left\{{\rm Tr}\left[H^{}_l H^{}_\nu G^{}_{l\nu}\right]\right\} = 0 \; ,\label{eq:I2} \\
%     (3)
{\cal I}^{}_3 &\equiv& {\rm Tr}\left\{ \left[G^{}_{l\nu}, H^{}_l \right]^3\right\} = 0 \; , \label{eq:I3}
%     (4)
\end{eqnarray}
where $H^{}_l \equiv M^{}_l M^\dagger_l$, $H^{}_\nu \equiv M^{}_\nu M^\dagger_\nu$ and $G^{}_{l\nu} \equiv M^{}_\nu H^*_l M^\dagger_\nu$ have been introduced, constitute a minimal set of sufficient and necessary conditions for CP conservation in the leptonic sector with massive Majorana neutrinos. It is worthwhile to emphasize that both charged-lepton and neutrino masses have been assumed to non-degenerate, which is actually true in light of recent neutrino oscillation data~\cite{Tanabashi:2018oca}. If neutrino masses were partially or completely degenerate, the number of sufficient and necessary conditions for CP conservation would be different~\cite{Branco:1986gr, Branco:1998bw, Mei:2003gu, Yu2019}.

On the other hand, it has been proved in Ref.~\cite{Branco:1986gr} that four vanishing WB invariants are indeed sufficient and necessary conditions for CP conservation in the leptonic sector with massive Majorana neutrinos, namely,
\begin{eqnarray}
\widehat{\cal I}^{}_1 &\equiv& {\rm Im}\left\{{\rm Tr}\left[H^{}_l H^{}_\nu G^{}_{l\nu}\right]\right\} = 0 \; ,
 \label{eq:Ip1} \\
%     (5)
\widehat{\cal I}^{}_2 &\equiv& {\rm Im}\left\{{\rm Tr}\left[H^{}_l H^2_\nu G^{}_{l\nu}\right]\right\} = 0 \; ,\label{eq:Ip2} \\
%     (6)
\widehat{\cal I}^{}_3 &\equiv& {\rm Im}\left\{{\rm Tr}\left[H^{}_l H^2_\nu G^{}_{l\nu} H^{}_\nu \right]\right\} = 0 \; , \label{eq:Ip3} \\
%     (7)
\widehat{\cal I}^{}_4 &\equiv& {\rm Im}\left\{{\rm Det}\left[G^{}_{l\nu} + H^{}_l H^{}_\nu \right]\right\} = 0 \; , \label{eq:Ip4}
%     (8)
\end{eqnarray}
where we have translated the original equations in Ref.~\cite{Branco:1986gr} into those in our notations defined in Eq.~(\ref{eq:lag}). As is well-known, there exist totally three CP-violating phases in the leptonic sector with Majorana neutrinos, one of which is of the Dirac type and the other two are of Majorana type. In the physical basis where the charged-lepton mass matrix $M^{}_l = \widehat{M}^{}_l \equiv {\rm Diag}\{m^{}_e, m^{}_\mu, m^{}_\tau\}$ and the Majorana neutrino mass matrix $M^{}_\nu = \widehat{M}^{}_\nu \equiv {\rm Diag}\{m^{}_1, m^{}_2, m^{}_3\}$ are both diagonal, the leptonic CP-violating phases will appear in the charged-current weak interaction through the flavor mixing matrix $V$, which is usually parametrized in terms of three mixing angles $\{\theta^{}_{12}, \theta^{}_{13}, \theta^{}_{23}\}$ and three CP-violating phases $\{\delta, \rho, \sigma\}$, i.e.,
\begin{eqnarray}\label{eq:parametrization}
V = \left( \begin{matrix} c^{}_{13} c^{}_{12} & c^{}_{13} s^{}_{12} & s^{}_{13} e^{-{\rm i}\delta} \cr -s_{12}^{} c_{23}^{} - c_{12}^{} s_{13}^{} s_{23}^{} e^{{\rm i}\delta}_{} & + c_{12}^{} c_{23}^{} - s_{12}^{} s_{13}^{} s_{23}^{} e^{{\rm i}\delta}_{} & c_{13}^{} s_{23}^{} \cr + s_{12}^{} s_{23}^{} - c_{12}^{} s_{13}^{} c_{23}^{} e^{{\rm i}\delta}_{} & - c_{12}^{} s_{23}^{} - s_{12}^{} s_{13}^{} c_{23}^{} e^{{\rm i}\delta}_{} & c_{13}^{} c_{23}^{} \end{matrix} \right) \cdot \left(\begin{matrix} e^{{\rm i}\rho} & 0 & 0 \cr 0 & e^{{\rm i}\sigma} & 0 \cr 0 & 0 & 1\end{matrix}\right) \; ,
\end{eqnarray}
 where $c^{}_{ij} \equiv \cos \theta^{}_{ij}$ and $s^{}_{ij} \equiv \sin \theta^{}_{ij}$ (for $ij = 12, 13, 23$) have been defined. Therefore, it is quite natural to claim that three vanishing WB invariants, as in Eqs.~(\ref{eq:I1})-(\ref{eq:I3}), should be the sufficient and necessary conditions for CP conservation.

In this paper, we raise the question whether the number of sufficient and necessary conditions for CP conservation in the leptonic sector with Majorana neutrinos is three or four. It can be easily verified that the WB invariant ${\cal I}^{}_1$ is proportional to the Jarlskog invariant ${\cal J}$~\cite{Jarlskog:1985ht, Wu:1985ea} in the leptonic sector, which can be explicitly calculated as ${\cal J} = s^{}_{12} c^{}_{12} s^{}_{23} c^{}_{23} s^{}_{13} c^2_{13} \sin\delta$ for the standard parametrization of $V$ in Eq.~(\ref{eq:parametrization}). Hence ${\cal I}^{}_1 = 0$ serves as the sufficient and necessary condition for a trivial Dirac CP-violating phase $\delta = 0$ or $180^\circ$, given the observed neutrino mixing angles $\theta^{}_{12} = 33.82^\circ$, $\theta^{}_{13} = 8.61^\circ$ and $\theta^{}_{23} = 48.3^\circ$~\cite{Esteban:2018azc}. Futhermore, the requirement for the other two {\it independent} WB invariants ${\cal I}^{}_2$ and ${\cal I}^{}_3$ to be vanishing gives rise to two {\it independent} equations for two Majorana CP-violating phases $\{\rho, \sigma\}$, which force these two phases to take only trivial values $0$ or $90^\circ$~\cite{Dreiner:2007yz}. However, as we shall explain later, this is true only when the yet unknown lightest neutrino mass $m^{}_1$ in the case of normal neutrino mass ordering turns out to be in a properly chosen range.\footnote{For clarity, we focus only on the case of normal neutrino mass ordering with $m^{}_1 < m^{}_2 < m^{}_3$. Nevertheless, the case of inverted neutrino mass ordering with $m^{}_3 < m^{}_1 < m^{}_2$ can be investigated in a similar way.}

The remaining part of our paper is structured as follows. In Sec.~\ref{sec:counter ex}, we give a concrete counter example, in which Eqs.~(\ref{eq:I1})-(\ref{eq:I3}) are satisfied while CP violation is still allowed. Then, we propose in Sec.~\ref{sec:new invariants} two new sets of three independent WB invariants. When they are vanishing, CP conservation is guaranteed for all the physical parameters within their experimentally allowed ranges. Finally, we summarize our main conclusions in Sec.~\ref{sec:summary}.

\section{A Counter Example}\label{sec:counter ex}

In this section, we give a concrete counter example, in which all the conditions in Eqs.~(\ref{eq:I1})-(\ref{eq:I3}) are fulfilled but CP violation is still present. Thus, one can conclude that Eqs.~(\ref{eq:I1})-(\ref{eq:I3}) cannot be the sufficient conditions for CP conservation in the leptonic sector with Majorana neutrinos. Since ${\cal I}^{}_i$ (for $i = 1, 2, 3$) are WB invariants, they can be explicitly calculated in any basis and the ultimate expressions should depend only on physical parameters, namely, the charged-lepton masses $\{m^{}_e, m^{}_\mu, m^{}_\tau\}$, neutrino masses $\{m^{}_1, m^{}_2, m^{}_3\}$, leptonic flavor mixing angles $\{\theta^{}_{12}, \theta^{}_{13}, \theta^{}_{23}\}$ and CP-violating phases $\{\delta, \rho, \sigma\}$. More explicitly, we can obtain
\begin{eqnarray}
{\cal I}_1 = - 6 {\rm i} \Delta^{}_{21} \Delta^{}_{31} \Delta^{}_{32} \Delta^{}_{e\mu} \Delta^{}_{\mu\tau} \Delta^{}_{\tau e} {\cal J} \; , \label{eq:I1expr}
%     (10)
\end{eqnarray}
where $\Delta^{}_{ij} \equiv m^2_i - m^2_j$ (for $i, j = 1, 2, 3$) and $\Delta^{}_{\alpha \beta} \equiv m^2_\alpha - m^2_\beta$ (for $\alpha, \beta = e, \mu, \tau$) denote the mass-squared differences for neutrinos and charged-leptons, respectively. In addition, the Jarlskog invariant ${\cal J} \equiv s^{}_{12} c^{}_{12} s^{}_{23} c^{}_{23} s^{}_{13} c^2_{13} \sin\delta  $ involves all the mixing angles $\{\theta^{}_{12}, \theta^{}_{13}, \theta^{}_{23}\}$ and the Dirac-type CP-violating phase $\delta$. Given non-degenerate neutrino masses with $\Delta^{}_{21} = 7.39\times 10^{-5}~{\rm eV}^2$ and $\Delta^{}_{31} = 2.523\times 10^{-3}~{\rm eV}^2$, as well as the observed mixing angles $\theta^{}_{12} = 33.82^\circ$, $\theta^{}_{13} = 8.61^\circ$ and $\theta^{}_{23} = 48.3^\circ$ from the latest global-fit analysis of current neutrino oscillation data~\cite{Esteban:2018azc}, ${\cal I}_1 = 0$ holds if and only if $\delta = 0$ or $180^\circ$. As a consequence, Eq.~(\ref{eq:I1}) ensures that the Dirac-type CP phase $\delta$ can take only trivial values. For simplicity, we shall assume $\delta = 0$ in the following discussions if ${\cal I}^{}_1 = 0$ is satisfied.

In order to prove that ${\cal I}^{}_i = 0$ (for $i = 1, 2, 3$) are not sufficient for CP conservation, we just need to give a counter example of nontrivial solutions to $\rho$ and $\sigma$. Once $\delta$ is forced to be zero by ${\cal I}^{}_1 = 0$, one can immediately observe that ${\cal I}^{}_2 = 0$ and ${\cal I}^{}_3 = 0$ lead to two {\it independent} identities for two Majorana-type CP phases, i.e.,
\begin{eqnarray}
0 &=& f^{}_1 \sin(2\rho) + f^{}_2 \sin(2\sigma) + f^{}_3 \sin(2\rho-2\sigma) \; ,\label{eq:I2expr} \\
%     (11)
0 &=& g^{}_1 \sin(2\rho) + g^{}_2 \sin(2\sigma) + g^{}_3 \sin(2\rho-2\sigma) \nonumber \\ &~& + g^{}_4 \sin(2\rho+2\sigma) + g^{}_5 \sin(2\rho-4\sigma) + g^{}_6 \sin(2\sigma-4\rho)
 \; , \label{eq:I3expr}
%    (12)
\end{eqnarray}
where $f^{}_i$ (for $i = 1, 2, 3$) and $g^{}_j$ (for $j = 1, 2, \cdots, 6$) are functions of three mixing angles and six lepton masses, but they are independent of $\rho$ and $\sigma$. The analytical expressions of these functions are rather lengthy, which are listed in the Appendix~\ref{appendixA} for reference. Although Eqs.~(\ref{eq:I2expr}) and (\ref{eq:I3expr}) are independent of each other, they are not \emph{linear} equations of $\rho$ and $\sigma$. Hence it is mathematically incorrect to claim that $\rho$ and $\sigma$ take only trivial values $0$ or $90^\circ$. For
some specific values of mixing angles, neutrino masses and charged-lepton masses, there indeed exist nontrivial solutions to $\rho$ and $\sigma$ in Eqs.~(\ref{eq:I2expr}) and (\ref{eq:I3expr}) such that the CP symmetry is violated.

Now we give a numerical example. First, we adopt the best-fit values of neutrino mass-squared differences $\Delta^{}_{21} = 7.39\times 10^{-5}~{\rm eV}^2$, $\Delta^{}_{31} = 2.523\times 10^{-3}~{\rm eV}^2$, and neutrino mixing angles $\theta^{}_{12} = 33.82^{\circ}$, $\theta^{}_{13} = 8.61^{\circ}$ and $\theta^{}_{23} = 48.3^{\circ}$ from Ref.~\cite{Esteban:2018azc}. Once the lightest neutrino mass $m^{}_1$ is known, we can determine the other two neutrino masses via $m^{}_2 = \sqrt{m^2_1 + \Delta^{}_{21}}$ and $m^{}_3 = \sqrt{m^2_1 + \Delta^{}_{31}}$. Moreover, the charged-lepton masses $m^{}_e = 0.511~{\rm MeV}$, $m^{}_{\mu} = 105.658~{\rm MeV}$ and $m^{}_{\tau} = 1776.86~{\rm MeV}$ have been precisely measured~\cite{Tanabashi:2018oca}. Then, one can see that the coefficients $f^{}_i$ (for $i = 1, 2, 3$) and $g^{}_j$ (for $j = 1, 2, \cdots, 6$) in Eqs.~(\ref{eq:I2expr}) and (\ref{eq:I3expr}) depend only on the unknown parameter $m^{}_1$. At present, the most restrictive bound on the lightest neutrino mass $m^{}_1$ comes from the precision measurements of the cosmic microwave background and the large-scale structures in our Universe~\cite{Aghanim:2018eyx}
\begin{eqnarray}\label{eq:planck}
m^{}_1 + m^{}_2 + m^{}_3 < 0.12~{\rm eV} \; ,
%     (13)
\end{eqnarray}
implying $0 \leq m^{}_1 < 0.04~{\rm eV}$ in the case of normal neutrino mass ordering. Taking $m^{}_1 = 0.03~{\rm eV}$ for example, together with the aforementioned values of other relevant parameters, we can rewrite Eqs.~(\ref{eq:I2expr}) and (\ref{eq:I3expr}) as
\begin{eqnarray}
{\cal I}_2^{\prime} &=& -2.092 \sin(2\rho) -8.754 \sin(2\sigma) - 0.035 \sin(2\rho-2\sigma) = 0 \; , \label{eq:I2new}\\
%     (14)
{\cal I}_3^{\prime} &=& 0.471 \sin(2\rho) - 3.535 \sin(2\sigma) + 1.177 \sin(2\rho-2\sigma) \nonumber \\
&~&  - 0.199 \sin(2\rho+2\sigma) - 2.574 \sin(2\rho-4\sigma) - 0.934 \sin(2\sigma-4\rho) = 0 \; , \label{eq:I3new}
%     (15)
\end{eqnarray}
where we have defined ${\cal I}_2^{\prime} \equiv {\cal I}^{}_2 /(10^6~{\rm eV}^4 \cdot {\rm MeV}^4)$ and ${\cal I}_3^{\prime} \equiv {\cal I}^{}_3 /(6{\rm i}\cdot 10^{24}~{\rm eV}^6 \cdot {\rm MeV}^{12})$ to make these two invariants dimensionless and to ensure that the numerical coefficients appearing in Eqs.~(\ref{eq:I2new}) and (\ref{eq:I3new}) are of ${\cal O}(1)$. These two equations have nontrivial solutions for $\rho$ and $\sigma$, i.e.,
\begin{eqnarray} \label{eq:sol}
\left\{
\begin{aligned}
\rho&=38.551^{\circ}\\
\sigma&=173.146^{\circ}\\
\end{aligned}
\right. \; ,
\qquad \text{or}\qquad
\left\{
\begin{aligned}
\rho&=141.449^{\circ}\\
\sigma&=6.854^{\circ}\\
\end{aligned}
\right. \; ,
%     (16)
\end{eqnarray}
which clearly indicates that CP violation is still present even when Eqs.~(\ref{eq:I1})-(\ref{eq:I3}) are satisfied. Note that according to the standard parametrization of the leptonic flavor mixing matrix $V$ in Eq.~(\ref{eq:parametrization}), the physical ranges of three CP-violating phases should be $\delta \in [0, 360^\circ)$ and $\rho, \sigma \in [0, 180^\circ)$.

%%%%%%%%%%%%%%%%%%%%%%%%%%%%%%%%%%% Fig. 1 %%%%%%%%%%%%%%%%%%%%%%%%%%%%%%
\begin{figure}[!t]
\centering
\includegraphics[width=0.7\textwidth]{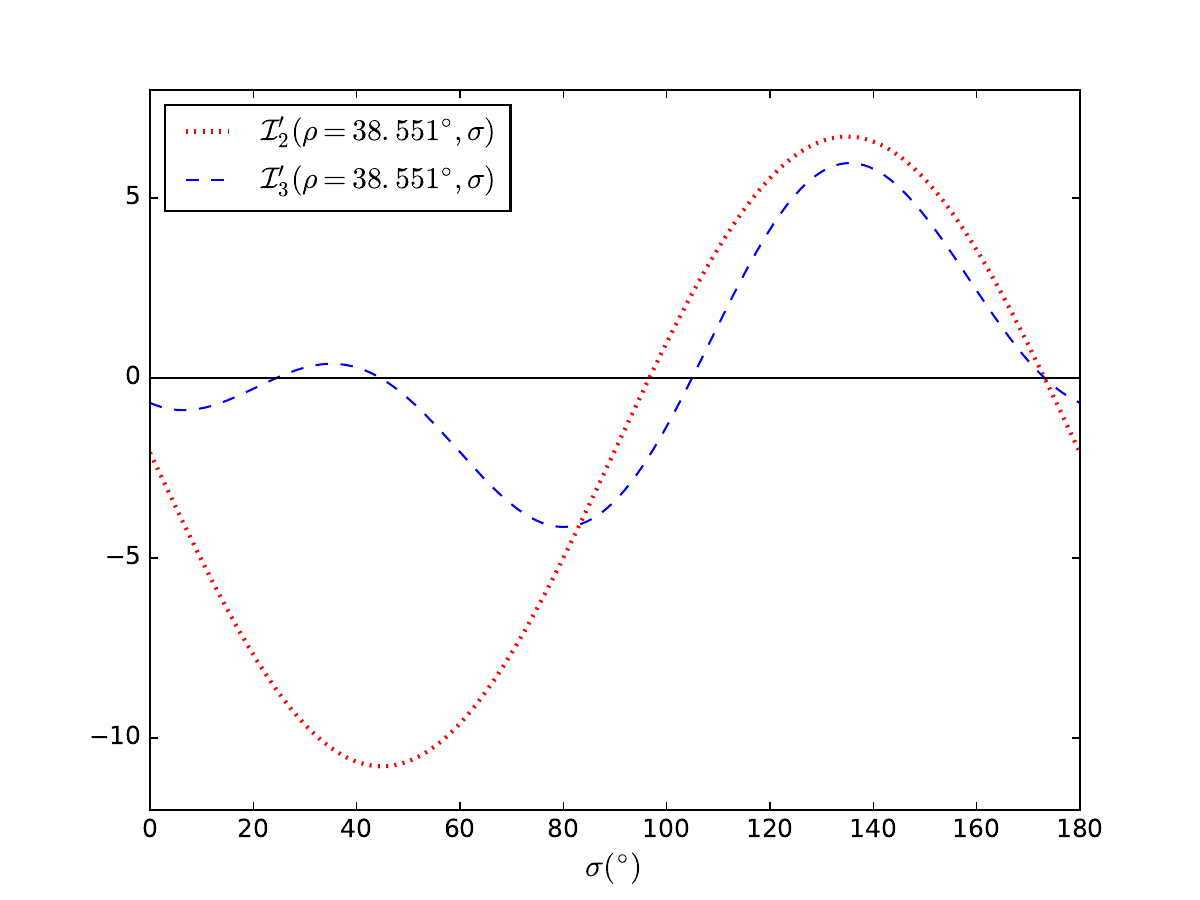}
\includegraphics[width=0.7\textwidth]{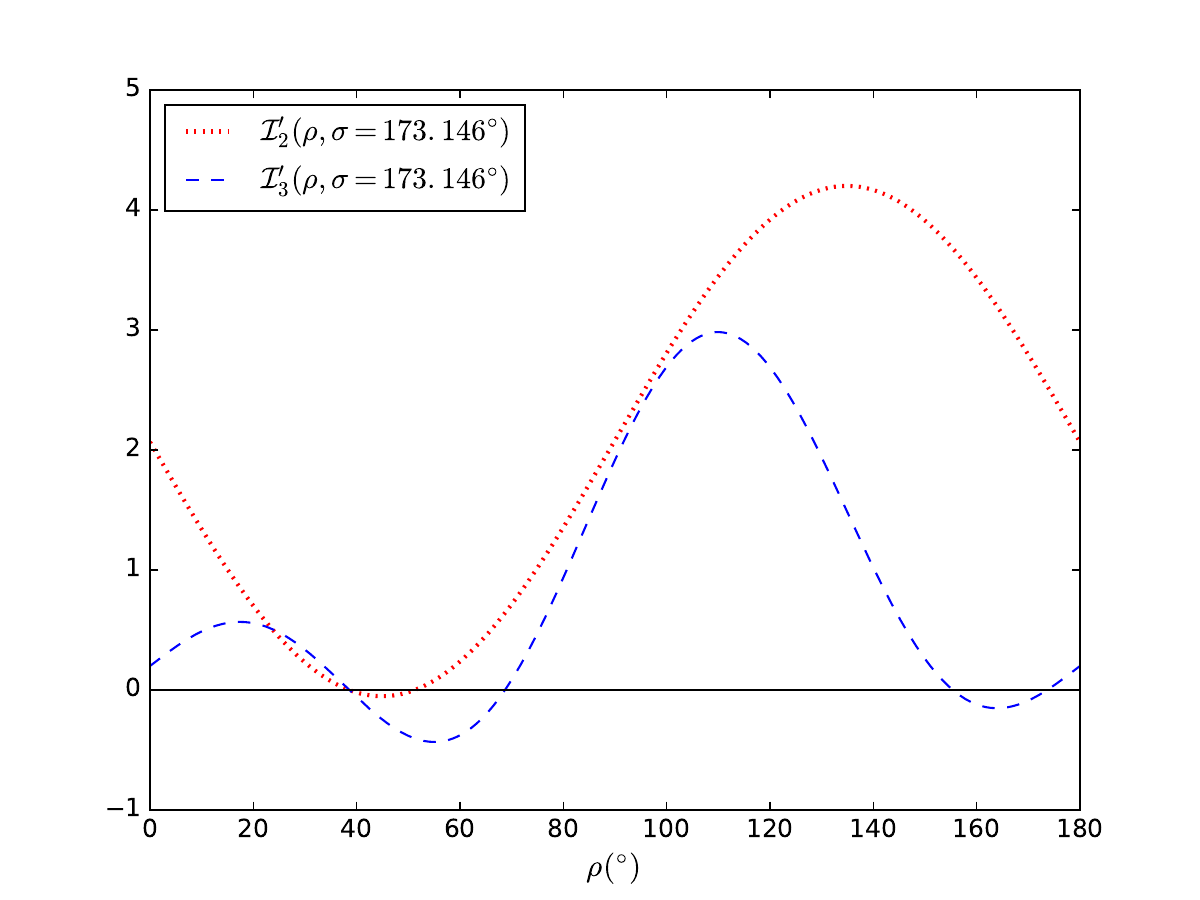}
\caption{Illustration for the dependence of two WB invariants ${\cal I}^\prime_2$ and ${\cal I}^\prime_3$ on the Majorana CP phases $\rho$ and $\sigma$. In the upper panel, we fix $\rho = 38.551^{\circ}$ and show the variation of ${\cal I}^\prime_2$ (red dotted curve) and ${\cal I}^\prime_3$ (blue dashed curve) against $\sigma$. In the lower panel, $\sigma=173.146^{\circ}$ is fixed and $\rho$ is varying in the range of $[0, 180^\circ)$. In both panels, two curves for ${\cal I}^\prime_2$ and ${\cal I}^\prime_3$ intersect at the common nontrivial zero point $(\rho, \sigma) = (38.551^\circ, 173.146^\circ)$.}\label{fig:Irhosigma}
\end{figure}
%%%%%%%%%%%%%%%%%%%%%%%%%%%%%%%%%%%%%%%%%%%%%%%%%%%%%%%%%%%%%%%%%%%%%%%%%%
As numerical calculations are always limited by their default precisions, one may wonder if those solutions in Eq.~(\ref{eq:sol}) are just numerical artifacts. To clarify this issue, we fix $\rho = 38.551^{\circ}$ and show how ${\cal I}_2^{\prime}$ and ${\cal I}_3^{\prime}$ change with respect to $\sigma$ in the upper panel of Fig.~\ref{fig:Irhosigma}. One can see that ${\cal I}^\prime_2$ (red dotted curve) and ${\cal I}^\prime_3$ (blue dashed curve) intersect and vanish at the common point $\sigma = 173.146^\circ$. As a double check, we fix $\sigma = 173.146^{\circ}$ and illustrate how ${\cal I}_2^{\prime}$ and ${\cal I}_3^{\prime}$ change with $\rho$ in the lower panel of Fig.~\ref{fig:Irhosigma}. The intersecting zero point of ${\cal I}_2^{\prime}$ and ${\cal I}_3^{\prime}$ is located at $\rho = 38.551^\circ$ as it should be. In both panels, significant deviations of ${\cal I}_2^{\prime}$ and ${\cal I}_3^{\prime}$ from zero can be observed when $\rho$ and $\sigma$ take the values other than their solutions in Eq.~(\ref{eq:sol}). The other set of solutions $(\rho, \sigma) = (141.449^\circ, 6.854^\circ)$ can be analyzed in a similar way. Therefore, it is quite convincing that the nontrivial solutions to $\rho$ and $\sigma$ are physically meaningful.

%%%%%%%%%%%%%%%%%%%%%%%%%%%%%%% Fig. 2 %%%%%%%%%%%%%%%%%%%%%%%%%%%%%%%%%%
\begin{figure}[!t]
\centering \includegraphics[width=0.7\textwidth]{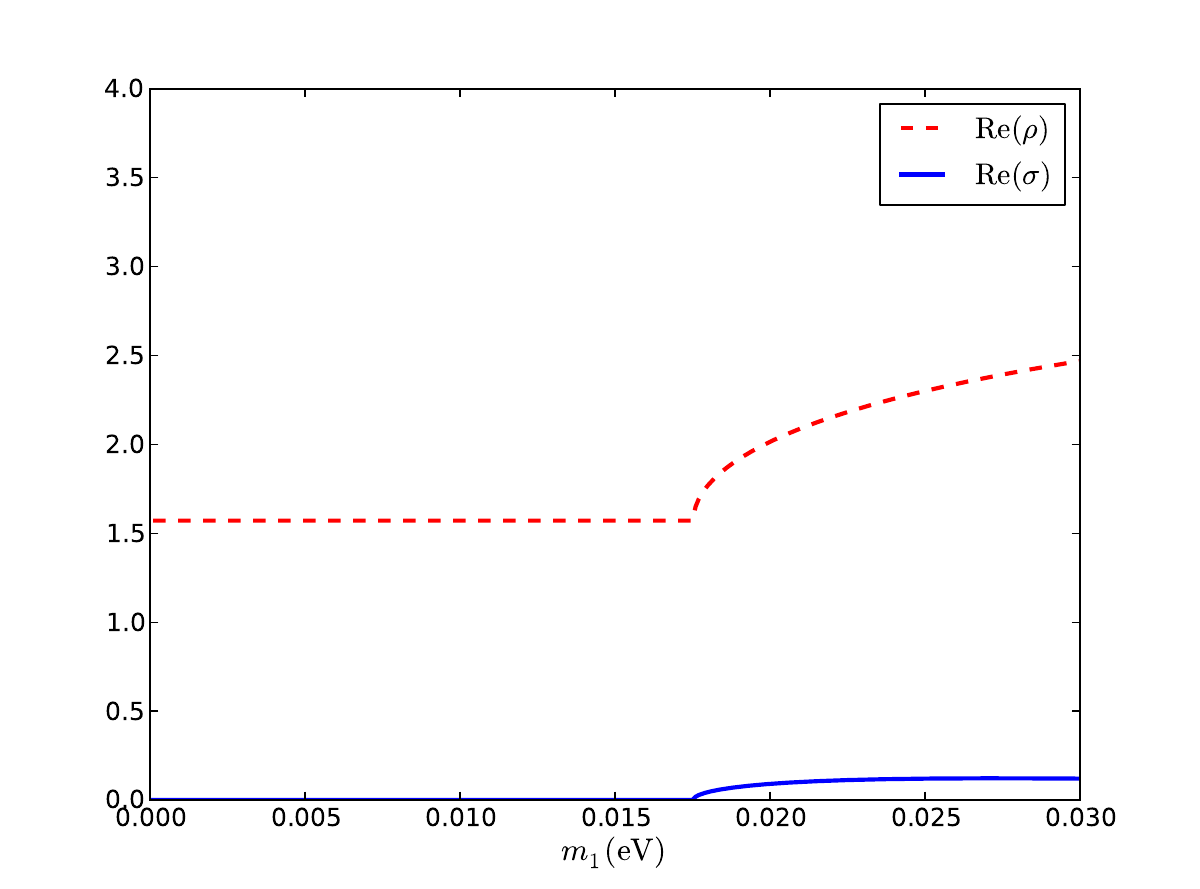}	\includegraphics[width=0.7\textwidth]{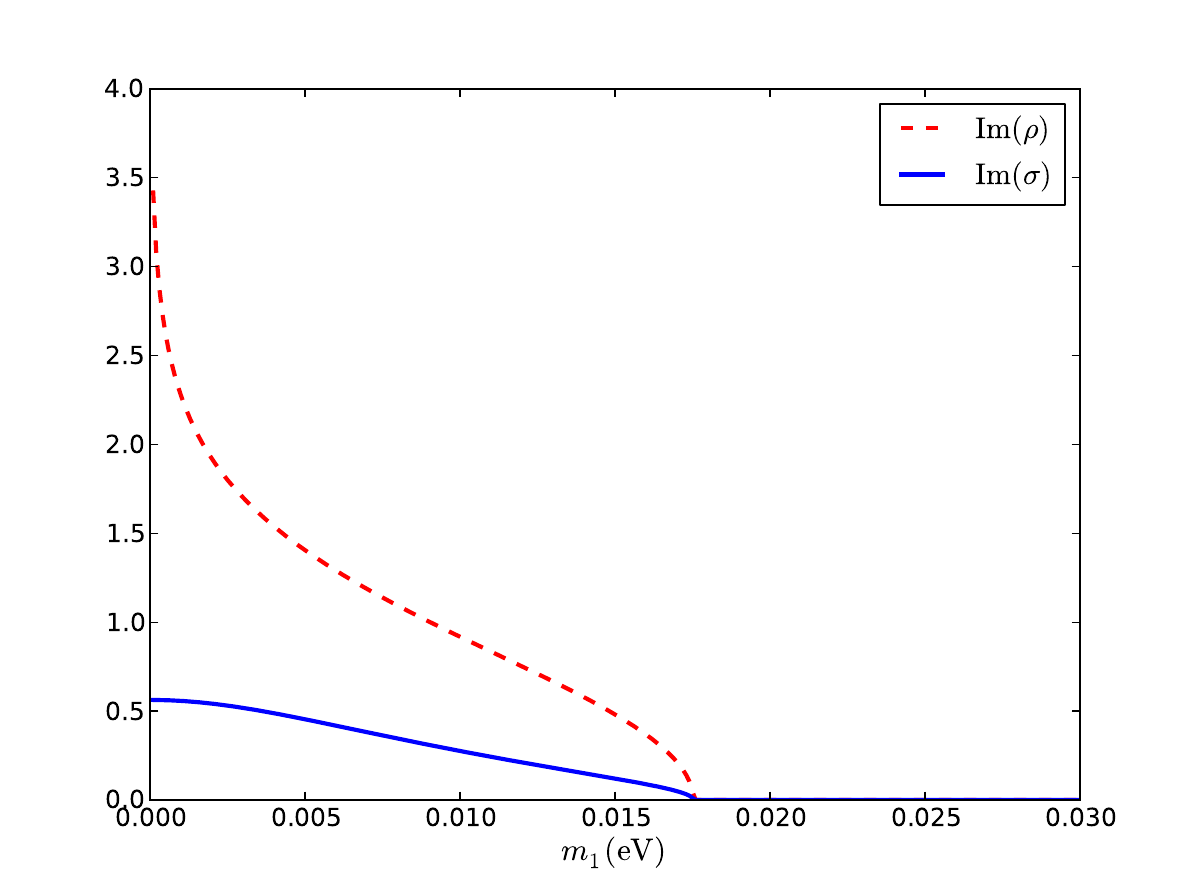}
\caption{Illustration for the real parts (the upper panel) and the imaginary parts (the lower panel) of the nontrivial solutions to $\rho$ (red dashed curve) and $\sigma$ (blue solid curve) for different values of the lightest neutrino mass $m^{}_1$.}\label{fig:realimag}
\end{figure}
%%%%%%%%%%%%%%%%%%%%%%%%%%%%%%%%%%%%%%%%%%%%%%%%%%%%%%%%%%%%%%%%%%%%%%%%%

We have demonstrated that for $m^{}_1 = 0.03~{\rm eV}$ there will be nontrivial CP-violating phases $\rho$ and $\sigma$ even when Eqs.~(\ref{eq:I1})-(\ref{eq:I3}) are satisfied. Hence one can conclude that ${\cal I}^{}_i = 0$ (for $i = 1, 2, 3$) are {\it not} the sufficient conditions for CP conservation in the leptonic sector with Majorana neutrinos. An immediate question is how the nontrivial solutions to $\rho$ and $\sigma$ depend on the lightest neutrino mass $m^{}_1$. To answer this question, we illustrate the variations of the real and imaginary parts of $\rho$ and $\sigma$ with respect to $m^{}_1$ in Fig.~\ref{fig:realimag}. The strategy to search for the nontrivial solutions is as follows. First, ${\cal I}^{}_1 = 0$ is required to ensure that $\delta = 0$ or $180^\circ$. Second, we numerically solve the equations of ${\cal I}^{}_2 = 0$ and ${\cal I}^{}_3 = 0$ for $\rho$ and $\sigma$, given an arbitrary value of $m^{}_1$ in the range of $(0, 0.04]~{\rm eV}$. Since $\rho$ and $\sigma$ should be real, the appearance of their imaginary parts implies that there are no meaningful nontrivial solutions. Interestingly, one can observe from both panels of Fig.~\ref{fig:realimag} that a critical value of $m^{}_*$ shows up at
\begin{eqnarray}\label{eq:critical}
m^{}_1 = m^{}_* \approx 0.0175~{\rm eV} \; ,
%     (17)
\end{eqnarray}
whose exact value certainly depends on the input values of other physical parameters. For $m^{}_1 > m^{}_*$, the nontrivial solutions of $\rho$ and $\sigma$ are real and thus physically allowed, so the CP conservation is absent. For $m^{}_1 \leq m^{}_*$, the imaginary parts of the solutions of $\rho$ and $\sigma$ become nonzero, which is physically meaningless, so the CP conservation is maintained. In the particular case of $m^{}_1 = 0$, as in the minimal version of type-I seesaw model for tiny Majorana neutrino masses~\cite{King:1999mb, Guo:2006qa}, since the Majorana CP phase $\rho$ associated with the mass eigenvalue $m^{}_1$ automatically disappears, we are left with only one physical Majorana CP phase $\sigma$. Therefore, either Eq.~(\ref{eq:I2}) or Eq.~(\ref{eq:I3}) is sufficient to guarantee CP conservation, which is well consistent with our numerical results in Fig.~\ref{fig:realimag}.

To conclude, when $0\leq m^{}_1 \leq m^{}_* \approx 0.0175~{\rm eV}$, Eqs.~(\ref{eq:I1})-(\ref{eq:I3}) guarantee CP conservation in the leptonic sector. But for $m^{}_1 > m^{}_* \approx 0.0175~{\rm eV}$, Eqs.~(\ref{eq:I1})-(\ref{eq:I3}) are not sufficient conditions for CP conservation. As the lightest neutrino mass is restricted into the range $0\leq m^{}_1 < 0.04~{\rm eV}$ by cosmological observations, we arrive at the final conclusion that Eqs.~(\ref{eq:I1})-(\ref{eq:I3}) are not sufficient and necessary conditions for CP conservation in the whole physically allowed parameter space.

\section{New Sets of Three WB Invariants}\label{sec:new invariants}
%%%%%%%%%%%%%%%%%%%%%%%%%%%%%%%%%%% Fig. 3 %%%%%%%%%%%%%%%%%%%%%%%%%%%%%%
\begin{figure}[!t]
\centering	\includegraphics[width=0.7\textwidth]{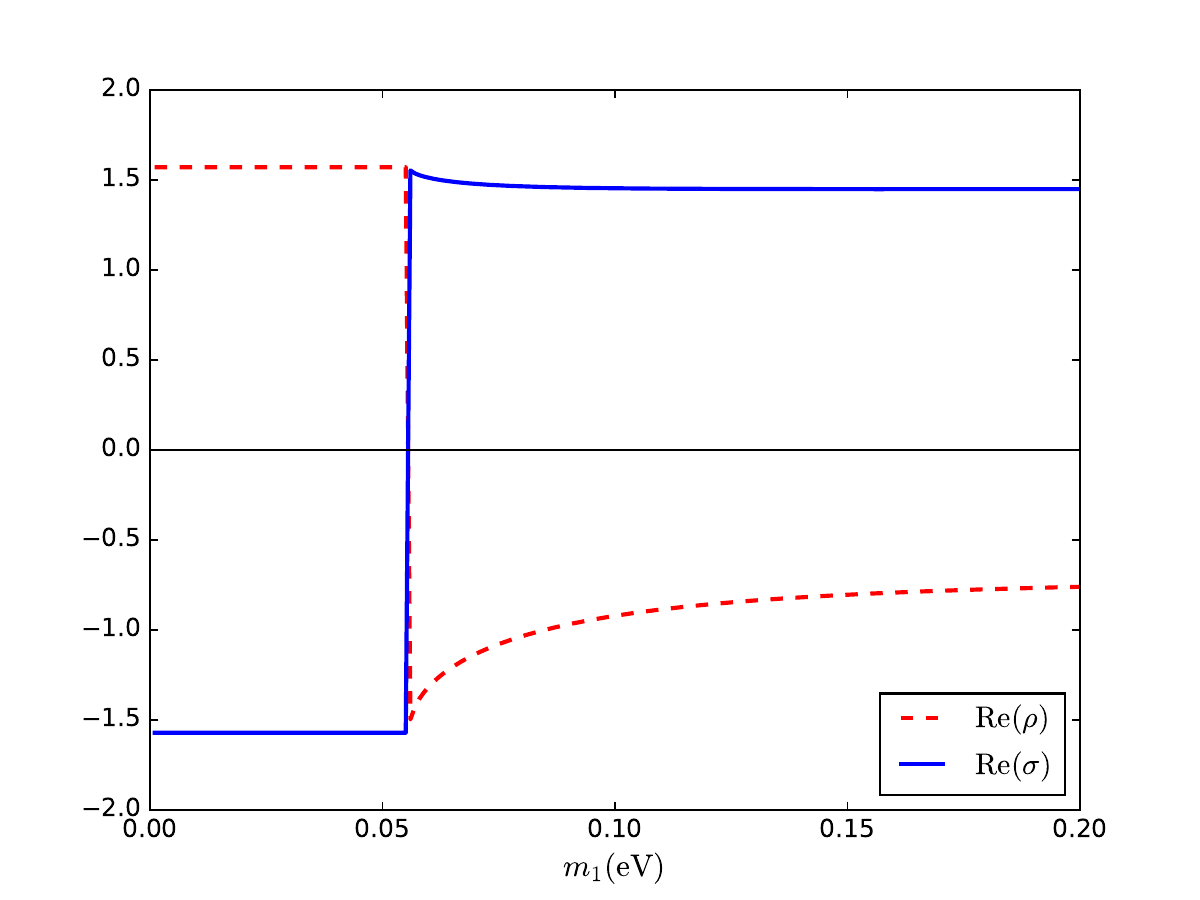}	\includegraphics[width=0.7\textwidth]{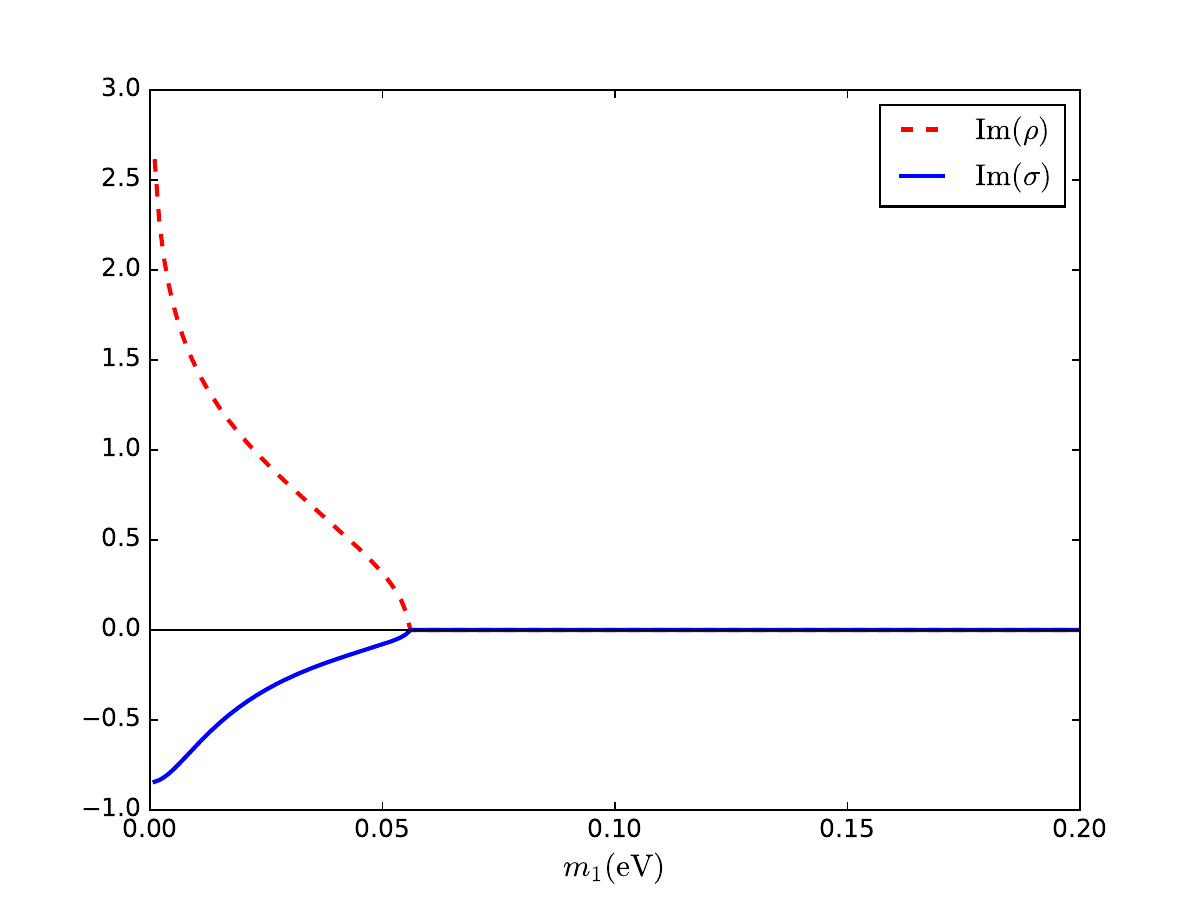}	\caption{Illustration for the real parts (the upper panel) and the imaginary parts (the lower panel) of the nontrivial solutions of $\rho$ (red dashed curve) and $\sigma$ (blue solid curve) for different values of $m^{}_1$, given the new set of three WB invariants $\{{\cal I}_1, {\cal I}_2, \widehat{\cal I}_2\}$.}\label{fig:realimag1}
\end{figure}
%%%%%%%%%%%%%%%%%%%%%%%%%%%%%%%%%%%%%%%%%%%%%%%%%%%%%%%%%%%%%%%%%%%%%%%%%

%%%%%%%%%%%%%%%%%%%%%%%%%%%%%%%%%% Fig. 4 %%%%%%%%%%%%%%%%%%%%%%%%%%%%%%%
\begin{figure}[!t]
	\centering	\includegraphics[width=0.7\textwidth]{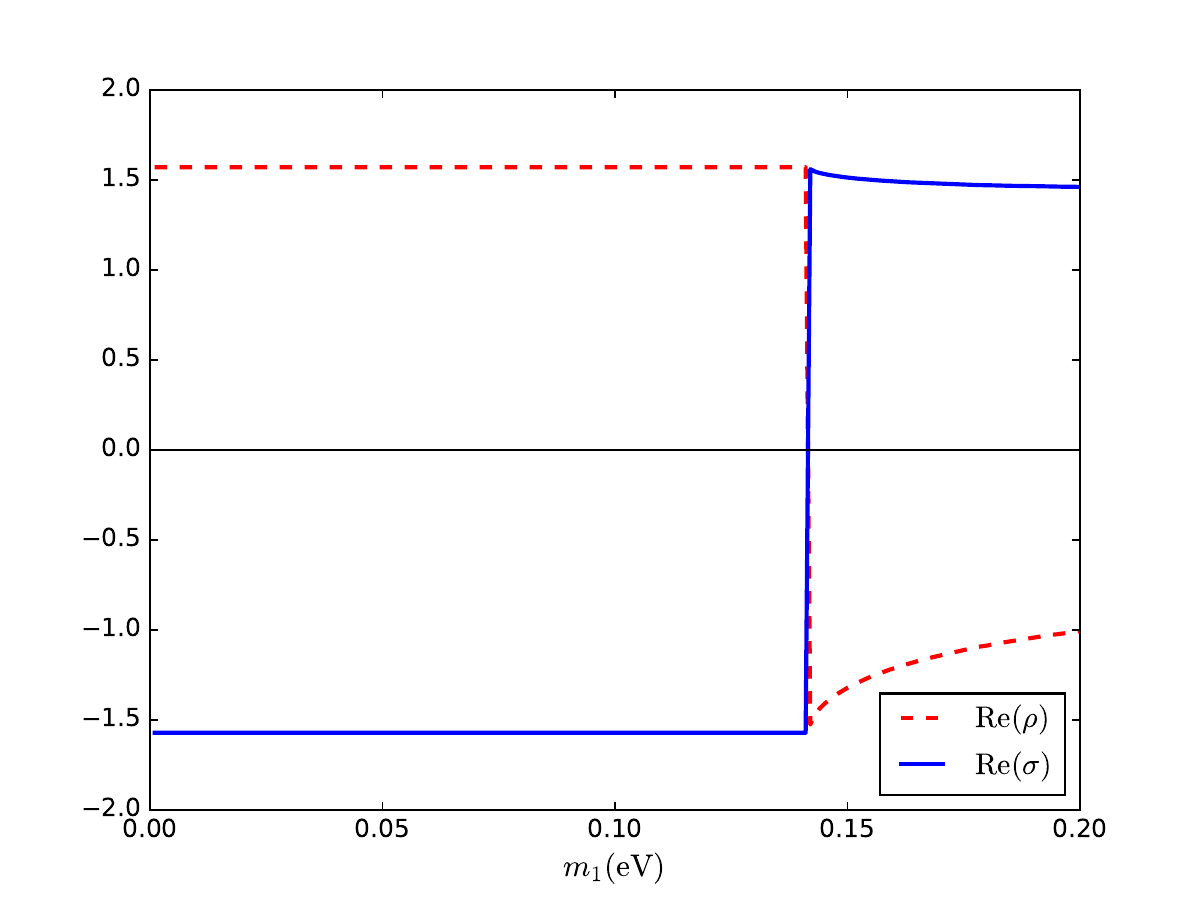}	\includegraphics[width=0.7\textwidth]{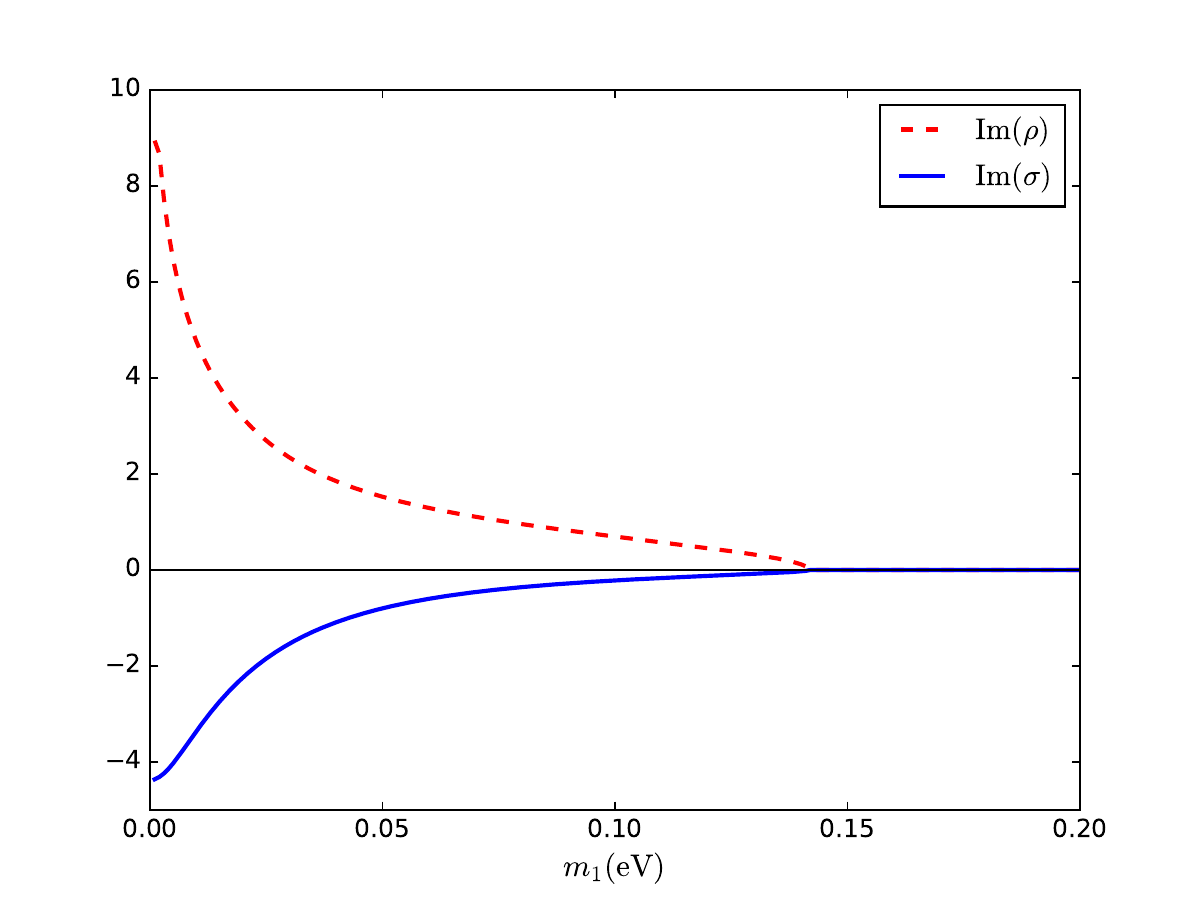}	\caption{Illustration for the real parts (the upper panel) and the imaginary parts (the lower panel) of the nontrivial solutions of $\rho$ (red dashed curve) and $\sigma$ (blue solid curve) for different values of $m^{}_1$, given the new set of three WB invariants $\{{\cal I}_1, \widehat{\cal I}_2, \widehat{\cal I}_3\}$.}\label{fig:realimag2}
\end{figure}
%%%%%%%%%%%%%%%%%%%%%%%%%%%%%%%%%%%%%%%%%%%%%%%%%%%%%%%%%%%%%%%%%%%%%%%%%%

Although the vanishing of three WB invariants in Eqs.~(\ref{eq:I1})-(\ref{eq:I3}) are not sufficient for CP conservation in the leptonic sector, it is intuitively expected that the number of sufficient and necessary conditions for CP conservation should be three, which is the total number of CP-violating phases in the flavor mixing matrix for massive Majorana neutrinos. On the other hand, one may be curious about what happens if another set of three WB invariants are chosen. Generally speaking, since the requirement for three independent WB invariants to be zero leads to three independent nonlinear equations of three CP phases, they are unable to enforce these phases to take just trivial values. If we choose another set of three WB invariants, then it is likely that there exists another critical value of $m^{}_1$, above which we can find real and nontrivial solutions for CP-violating phases. However, different sets of three WB invariants give rise to different critical values for $m^{}_1$. Therefore, what we need to do is to find out a new set of three WB invariants, which render the critical value of $m^{}_1$ to be larger than the upper bound $m^{}_1 < 0.04~{\rm eV}$ such that CP conservation is guaranteed at least for all the physical parameters within the experimentally allowed regions.

For this purpose, we retain the WB invariant ${\cal I}^{}_1$ so that ${\cal I}^{}_1 = 0$ forces the Dirac CP phase $\delta$ to be $0$ or $180^\circ$. As has been proved in Ref.~\cite{Branco:1986gr}, it is always possible to construct a series of WB invariants in the form of ${\cal I}^{kmn}_{rst} \equiv {\rm Im}\left\{{\rm Tr}\left[H^{k}_l H^{m}_\nu G^{n}_{l\nu} H^{r}_l H^{s}_\nu G^{t}_{l\nu}\cdots\right]\right\}$, where the nonnegative integers $\{k, m, n, r, s, t\}$ are the power indices and ``$\cdots$" denote the additional matrices formed of $H^{}_l$, $H^{}_\nu$ and $G^{}_{l\nu}$. It is straightforward to verify these quantities ${\cal I}^{kmn}_{rst}$ are invariant under the WB transformations and they are related to all three CP-violating phases in the flavor mixing matrix.

 For illustration, we propose a new set of three WB invariants $\{{\cal I}^{}_1, {\cal I}^{}_2, \widehat{\cal I}^{}_2\}$, where $\widehat{\cal I}^{}_2$ has been given in Eq.~(\ref{eq:Ip2}). Under the condition ${\cal I}^{}_1 = 0$, which requires $\delta = 0$, one can explicitly calculate ${\cal I}^{}_2 = 0$ and $\widehat{\cal I}^{}_2 = 0$. While the former leads to the same identity in Eq.~(\ref{eq:I2expr}), the latter gives
\begin{eqnarray}
0 = h^{}_1 \sin(2\rho) + h^{}_2 \sin(2\sigma) + h^{}_3 \sin(2\rho-2\sigma) \; ,\label{eq:Ip2expr}
%     (18)
\end{eqnarray}
where $h^{}_i$ (for $i = 1, 2, 3$) are functions of charged-lepton masses, neutrino masses and flavor mixing angles. The explicit expressions of $h^{}_i$ (for $i = 1, 2, 3$) are also summarized in the Appendix~\ref{appendixA}. Solving Eqs.~(\ref{eq:I2expr}) and (\ref{eq:Ip2expr}) for $\rho$ and $\sigma$ in the same manner as before, we indeed find nontrivial solutions if $m^{}_1$ is larger than the critical value
\begin{eqnarray}\label{eq:critical1}
m^{}_1 = m_*^{\prime} \approx 0.0557~{\rm eV} \; ,
%     (19)
\end{eqnarray}
where the same input values of other physical parameters as in Eq.~(\ref{eq:critical}) are taken. In Fig.~\ref{fig:realimag1}, we have shown the real and imaginary parts of $\rho$ and $\sigma$ against the lightest neutrino mass $m^{}_1$, where the critical point at $m^{}_1 = m^\prime_* \approx 0.0557~{\rm eV}$ can be easily identified. In addition, one can observe that for $m^{}_1 \leq m^\prime_*$, the Majorana CP phases $\rho$ and $\sigma$ have both real and imaginary parts. As we have mentioned, it is meaningless for $\rho$ and $\sigma$ to be complex. Therefore, these nontrivial solutions to $\rho$ and $\sigma$ are not physical. For $m^{}_1 > m^\prime_*$, the imaginary parts of $\rho$ and $\sigma$ vanish, and the real parts deviate from the trivial value of $\rho = 90^\circ$ and $\sigma = -90^\circ$. This is a clear indication of nontrivial and physical solutions of $\rho$ and $\sigma$ to the equations of ${\cal I}^{}_2 = 0$ and $\widehat{\cal I}^{}_2 = 0$. However, this happens only for $m^{}_1 > m^\prime_* \approx 0.0557~{\rm eV}$, which turns out to be in contradiction with the cosmological bound on neutrino masses in Eq.~(\ref{eq:planck}). Therefore, it is reasonable to claim that the vanishing of all the new set of three WB invariants $\{{\cal I}^{}_1, {\cal I}^{}_2, \widehat{\cal I}^{}_2\}$ constitutes sufficient and necessary conditions for CP conservation in the leptonic sector with massive Majorana neutrinos when all the physical parameters are lying within their experimentally allowed regions.

It should be noted that our choice of WB invariants is by no means unique. It is possible to find another set of three WB invariants to guarantee CP conservation, as long as the corresponding critical value of $m^{}_1$ is larger than its cosmological upper bound $0.04~{\rm eV}$. For instance, we can also take $\{{\cal I}_1, \widehat{\cal I}^{}_2, \widehat{\cal I}^{}_3\}$ as the alternative set of WB invariants. The implications of ${\cal I}^{}_1 = 0$ and $\widehat{\cal I}^{}_2 = 0$ have already been discussed, while the fulfillment of $\widehat{\cal I}^{}_3 = 0$ implies the following equation
\begin{eqnarray}\label{eq:Ip3expr}
0 = k^{}_1 \sin(2\rho) + k^{}_2 \sin(2\sigma) + k^{}_3 \sin(2\rho - 2\sigma) \; ,
%     (20)
\end{eqnarray}
where the explicit expressions of $k^{}_i$ (for $i = 1, 2, 3$) can be found in the Appendix~\ref{appendixA}. For this set of WB invariants, the critical value of $m^{}_1$ is found to be
\begin{eqnarray}\label{eq:critical2}
m^{}_1 = m_*^{\prime\prime} \approx 0.142~{\rm eV} \; ,
%     (21)
\end{eqnarray}
which is even larger than the cosmological upper bound $0.12~{\rm eV}$ on the sum of three neutrino masses. Following the same approach as before, we first obtain the solution of $\delta = 0$ to ${\cal I}^{}_1 = 0$, and then calculate the real and imaginary parts of the solutions of $\rho$ and $\sigma$ to the equations $\widehat{\cal I}^{}_2 = 0$ and $\widehat{\cal I}^{}_3 = 0$. The numerical results are presented in Fig.~\ref{fig:realimag2}, where the critical value of $m^{}_1 = m^{\prime\prime}_* \approx 0.142~{\rm eV}$ can be well recognized.

Finally, we stress that although any set of three WB invariants cannot {\it in general} guarantee CP conservation, a set of four WB invariants are sufficient to achieve this goal as first suggested in Ref.~\cite{Branco:1986gr}. To be specific, we consider a set of four WB invariants $\{{\cal I}^{}_1, {\cal I}^{}_2, \widehat{\cal I}^{}_2, \widehat{\cal I}^{}_3\}$. Then it is straightforward to prove that the vanishing of all these invariants is the sufficient and necessary conditions for CP conservation in the leptonic sector with massive Majorana neutrinos. The proof is as follows. First, ${\cal I}^{}_1 = 0$ is equivalent to $\delta = 0$ or $180^\circ$, as we have known from the previous discussions. Next, we notice that ${\cal I}^{}_2 = 0$, $\widehat{\cal I}^{}_2 = 0$ and $\widehat{\cal I}^{}_3 = 0$ lead to Eqs.~(\ref{eq:I2expr}), (\ref{eq:Ip2expr}) and (\ref{eq:Ip3expr}), respectively. These equations can be recast into the matrix form
\begin{eqnarray}\label{eq:matrix}
\left(\begin{matrix} f^{}_1 & f^{}_2 & f^{}_3 \cr h^{}_1 & h^{}_2 & h^{}_3 \cr k^{}_1 & k^{}_2 & k^{}_3\end{matrix}\right) \cdot \left(\begin{matrix} \sin(2\rho) \cr \sin(2\sigma) \cr \sin(2\rho - 2\sigma)\end{matrix}\right) = {\bf 0} \; .
%     (22)
\end{eqnarray}
The determinant of the coefficient matrix, denoted as ${\cal A}$, on the left-hand side of Eq.~(\ref{eq:matrix}) reads
\begin{eqnarray} \label{eq:det}
{\rm Det}({\cal A}) = h^2_{12} h^2_{13} h^2_{23} m^2_1 m^2_2 m^2_3 \Delta^2_{21} \Delta^2_{31} \Delta^2_{32} \; ,
%     (23)
\end{eqnarray}
where $h^{}_{ij} \equiv |\left(H^{}_l\right)^{}_{ij}|$ (for $ij = 12, 13, 23$) have been defined. Note that $H^{}_l$ here should be evaluated in the basis where $M^{}_\nu$ is diagonal and in the assumption of $\delta = 0$, as we explain in the Appendix~\ref{appendixA}. Given the experimentally observed charged-lepton masses and neutrino mixing angles, one can verify that ${\rm Det}({\cal A}) \neq 0$.\footnote{An exceptional case is $m^{}_1 = 0$. However, as we have mentioned in Sec.~\ref{sec:counter ex}, only the Majorana CP phase $\sigma$ is left in this case. So it will be forced to take trivial values if any one of $\{{\cal I}^{}_2, \widehat{\cal I}^{}_2, \widehat{\cal I}^{}_3 \}$ vanishes.} Consequently, only zero solutions exist for the variables $\sin(2\rho)$, $\sin(2\sigma)$ and $\sin(2\rho - 2\sigma)$ in Eq.~(\ref{eq:matrix}). This completes the proof of CP conservation. It is worthwhile to mention that this proof is valid no matter what value $m^{}_1$ may take.

\section{Summary}\label{sec:summary}

In this paper, we raise the question whether the number of sufficient and necessary conditions for CP conservation in the leptonic sector with massive Majorana neutrinos is three or four. The final answer to this question can be summarized as below
\begin{itemize}
\item Four conditions, such as those in Eqs.~(\ref{eq:Ip1})-(\ref{eq:Ip4}) and ${\cal I}^{}_1 = {\cal I}^{}_2 = \widehat{\cal I}^{}_2 = \widehat{\cal I}^{}_3 = 0$, are sufficient and necessary for CP conservation, which is independent of the yet-unknown lightest neutrino mass $m^{}_1$. However, the number of sufficient and necessary conditions is larger than that of CP-violating phases.

\item Three conditions, such as ${\cal I}^{}_1 = {\cal I}^{}_2 = \widehat{\cal I}^{}_2 = 0$ and ${\cal I}^{}_1 = \widehat{\cal I}^{}_2 = \widehat{\cal I}^{}_3 = 0$, are sufficient and necessary for CP conservation, in the assumption that $m^{}_1 < m^\prime_* \approx 0.0557~{\rm eV}$ in the former case and $m^{}_1 < m^{\prime\prime}_* \approx 0.142~{\rm eV}$ in the latter case.
\end{itemize}

Although we concentrate only on the case of normal neutrino mass ordering, it is quite obvious that our analysis can be extended to the case of inverted neutrino mass ordering. Moreover, in the scenarios of three sufficient conditions, the critical values of $m^{}_1$ depend very much on the choice of WB invariants as well as the input values of other physical parameters.

We have not attempted to perform a systematic study of sufficient and necessary conditions for leptonic CP conservation, which will be the topic of another work~\cite{Yu:2020gre}. However, by giving a concrete counter example, we have demonstrated that even if three conditions in Eqs.~(\ref{eq:I1})-(\ref{eq:I3}) are satisfied, there will be CP violation in the leptonic sector for $m^{}_1 > m^{}_* \approx 0.0175~{\rm eV}$. Given the observationally allowed region $m^{}_1 < 0.04~{\rm eV}$, two new sets of three WB invariants, namely, $\{{\cal I}^{}_1, {\cal I}^{}_2, \widehat{\cal I}^{}_2\}$ and $\{{\cal I}^{}_1, \widehat{\cal I}^{}_2, \widehat{\cal I}^{}_3\}$, have been proposed such that ${\cal I}^{}_1 = {\cal I}^{}_2 = \widehat{\cal I}^{}_2 = 0$ or ${\cal I}^{}_1 = \widehat{\cal I}^{}_2 = \widehat{\cal I}^{}_3 = 0$ serves as sufficient and necessary conditions for CP conservation. Such an investigation should be very suggestive for our understanding of leptonic CP violation~\cite{Branco:2011zb}, which is the primary task for the future neutrino oscillation experiments.

\section*{Acknowledgements}

This work was supported in part by the National Natural Science Foundation of China under Grant No.~11775232 and No.~11835013, and by the CAS Center for Excellence in Particle Physics.

\begin{appendix}

\section{Analytical Expressions of Relevant Functions}\label{appendixA}

In this appendix, we collect the explicit expressions of all the relevant functions appearing in the WB invariants, which are constructed via three Hermitian matices $H^{}_l \equiv M^{}_l M^\dagger_l$, $H^{}_\nu = M^{}_\nu M^\dagger_\nu$ and $G^{}_{l\nu} \equiv M^{}_\nu H^*_l M^\dagger_\nu$. By definition, the WB invariants are independent of the flavor basis in which they are calculated. Therefore, we work in the basis where the Majorana neutrino mass matrix $M^{}_\nu = \widehat{M}^{}_\nu \equiv {\rm Diag}\{m^{}_1, m^{}_2, m^{}_3\}$ is diagonal. In this case, we have $H^{}_\nu = {\rm Diag}\{m^2_1, m^2_2, m^2_3\}$ and
\begin{eqnarray}
\left(H^{}_l\right)^{}_{ij} &=& \sum_{\alpha = e, \mu, \tau} m^2_\alpha V^*_{\alpha i} V^{}_{\alpha j} \; , \label{eq:Hl} \\
%     (24)
\left(G^{}_{l\nu}\right)^{}_{ij} &=& \sum_{\alpha = e, \mu, \tau} m^{}_i m^{}_j m^2_\alpha V^{}_{\alpha i} V^*_{\alpha j} \; , \label{eq:Glnu}
%     (25)
\end{eqnarray}
where $V$ is the leptonic flavor mixing matrix given in Eq.~(\ref{eq:parametrization}). For every set of WB invariants in question, we have chosen ${\cal I}^{}_1$ as one element such that ${\cal I}^{}_1 = 0$ ensures $\delta = 0$ or $180^\circ$. Under this condition, Eqs.~(\ref{eq:Hl}) and (\ref{eq:Glnu}) will be further simplified to
\begin{eqnarray}
\left(H^{}_l\right)^{}_{ij} &=& \sum_{\alpha = e, \mu, \tau} m^2_\alpha U^{}_{\alpha i} U^{}_{\alpha j} e^{-{\rm i}\omega^{}_{ij}}\; , \label{eq:Hlsimp} \\
%     (26)
\left(G^{}_{l\nu}\right)^{}_{ij} &=& \sum_{\alpha = e, \mu, \tau} m^{}_i m^{}_j m^2_\alpha U^{}_{\alpha i} U^{}_{\alpha j} e^{+{\rm i}\omega^{}_{ij}} \; , \label{eq:Glnusimp}
%     (27)
\end{eqnarray}
where the nonzero phases are $\omega^{}_{12} = -\omega^{}_{21} = \rho - \sigma$, $\omega^{}_{13} = - \omega^{}_{31} = \rho$ and $\omega^{}_{23} = -\omega^{}_{32} = \sigma$, and $U$ denotes the real and orthogonal matrix obtained by setting all three phases in the mixing matrix $V$ to be zero, namely,
\begin{eqnarray}\label{eq:U}
U = \left( \begin{matrix} c^{}_{13} c^{}_{12} & c^{}_{13} s^{}_{12} & s^{}_{13} \cr -s_{12}^{} c_{23}^{} - c_{12}^{} s_{13}^{} s_{23}^{} & + c_{12}^{} c_{23}^{} - s_{12}^{} s_{13}^{} s_{23}^{} & c_{13}^{} s_{23}^{} \cr + s_{12}^{} s_{23}^{} - c_{12}^{} s_{13}^{} c_{23}^{} & - c_{12}^{} s_{23}^{} - s_{12}^{} s_{13}^{} c_{23}^{} & c_{13}^{} c_{23}^{} \end{matrix} \right) \; ,
%    (28)
\end{eqnarray}
where only three mixing angles $\{\theta^{}_{12}, \theta^{}_{13}, \theta^{}_{23}\}$ are involved, and $c^{}_{ij} \equiv \cos \theta^{}_{ij}$ and $s^{}_{ij} \equiv \sin \theta^{}_{ij}$ have been defined.

Now it is straightforward to calculate the elements $h^{}_{ij} \equiv \sum_\alpha^{} m_\alpha^2 U_{\alpha i}^{}U_{\alpha j}^{}$ of $H^{}_l$ in Eq.~(\ref{eq:Hlsimp}). For the off-diagonal elements, we obtain
\begin{eqnarray}
h^{}_{12} &=& \sum_{\alpha = e, \mu, \tau} m^2_\alpha U^{}_{\alpha 1} U^{}_{\alpha 2} = U^{}_{\mu 1} U^{}_{\mu 2} \Delta^{}_{\mu e} + U^{}_{\tau 1} U^{}_{\tau 2} \Delta^{}_{\tau e} \; , \nonumber \\
h^{}_{13} &=& \sum_{\alpha = e, \mu, \tau} m^2_\alpha U^{}_{\alpha 1} U^{}_{\alpha 3} = U^{}_{\mu 1} U^{}_{\mu 3} \Delta^{}_{\mu e} + U^{}_{\tau 1} U^{}_{\tau 3} \Delta^{}_{\tau e} \; , \label{eq:hij}\\
h^{}_{23} &=& \sum_{\alpha = e, \mu, \tau} m^2_\alpha U^{}_{\alpha 2} U^{}_{\alpha 3} = U^{}_{\mu 2} U^{}_{\mu 3} \Delta^{}_{\mu e} + U^{}_{\tau 2} U^{}_{\tau 3} \Delta^{}_{\tau e} \; , \nonumber
\end{eqnarray}
where the orthogonality conditions $U^{}_{ei} U^{}_{ej} + U^{}_{ei} U^{}_{ej} + U^{}_{ei} U^{}_{ej} = 0$ for $ij = 12, 13, 23$ have been used and $\Delta^{}_{\alpha \beta} \equiv m^2_\alpha - m^2_\beta$ for $\alpha, \beta = e, \mu, \tau$ have been defined. For the diagonal elements, one gets
\begin{eqnarray}
h^{}_{11} &=& \sum_{\alpha = e, \mu, \tau} m^2_\alpha U^2_{\alpha 1} = m^2_e + U^2_{\mu 1} \Delta^{}_{\mu e} + U^2_{\tau 1} \Delta^{}_{\tau e} \; , \nonumber \\
h^{}_{22} &=& \sum_{\alpha = e, \mu, \tau} m^2_\alpha U^2_{\alpha 2} = m^2_\mu +  U^2_{e 2} \Delta^{}_{e \mu} + U^2_{\tau 2} \Delta^{}_{\tau \mu} \; , \label{eq:hii}\\
h^{}_{33} &=& \sum_{\alpha = e, \mu, \tau} m^2_\alpha U^2_{\alpha 3} = m^2_\tau + U^2_{e 3} \Delta^{}_{e \tau} + U^2_{\mu 3} \Delta^{}_{\mu \tau} \; , \nonumber
\end{eqnarray}
where the normalization conditions $U^2_{ei} + U^2_{\mu i} + U^2_{\tau i} = 1$ for $i = 1, 2, 3$ have been implemented. According to the parametrization of $U$ in Eq.~(\ref{eq:U}), we finally arrive at
\begin{eqnarray}
h^{}_{12} &=& - s^{}_{12}c^{}_{12}c^2_{13} \Delta^{}_{\mu e} + \left[s^{}_{12}c^{}_{12} (s^2_{13} c^2_{23} - s^2_{23}) + (c^2_{12} - s^2_{12})s^{}_{13} s^{}_{23} c^{}_{23} \right] \Delta^{}_{\tau \mu} \;, \nonumber \\
h^{}_{13} &=& -c^{}_{12} s^{}_{13} c^{}_{13} \Delta^{}_{\mu e} + (s^{}_{12} s^{}_{23} - c^{}_{12} s^{}_{13} c^{}_{23})c^{}_{13} c^{}_{23} \Delta^{}_{\tau \mu} \; , \nonumber \\
h^{}_{23} &=& -s^{}_{12} s^{}_{13} c^{}_{13} \Delta^{}_{\mu e} - (c^{}_{12} s^{}_{23} + s^{}_{12} s^{}_{13} c^{}_{23})c^{}_{13} c^{}_{23} \Delta^{}_{\tau \mu} \; , \nonumber \\
h^{}_{11} &=& m^2_e + (1 - c^2_{12} c^2_{13})\Delta^{}_{\mu e} + (s^{}_{12} s^{}_{23} - c^{}_{12} s^{}_{13} c^{}_{23})^2 \Delta^{}_{\tau\mu} \; , \nonumber \\
h^{}_{22} &=& m^2_\mu - s^2_{12}c^2_{13}\Delta^{}_{\mu e} + (c_{12}^{} s_{23}^{} + s_{12}^{} s_{13}^{} c_{23}^{})^2 \Delta^{}_{\tau\mu} \; , \nonumber \\
h^{}_{33} &=& m^2_\tau - s^2_{13} \Delta^{}_{\mu e} - (s^2_{13} + c^2_{13} s^2_{23})\Delta^{}_{\tau \mu} \; , \nonumber
\end{eqnarray}
where we have retained only two independent mass-squared differences $\Delta^{}_{\mu e}$ and $\Delta^{}_{\tau \mu}$ for charged leptons. From the last three equations, one can verify that $h^{}_{11} + h^{}_{22} + h^{}_{33} = m^2_e + m^2_\mu + m^2_\tau$ holds.

With the help of Eqs.~(\ref{eq:hij}) and (\ref{eq:hii}), we can explicitly compute the WB invariants ${\cal I}^{}_2$, $\widehat{\cal I}^{}_2$ and $\widehat{\cal I}^{}_3$ and extract the relevant functions $f^{}_i$, $h^{}_i$ and $k^{}_i$ (for $i = 1, 2, 3$). The final results are relatively simple and can be grouped into the matrix ${\cal A}$ that has been introduced in Eq.~(\ref{eq:matrix}), viz.,
\begin{eqnarray}\label{eq:Amatrix}
{\cal A} = \left(\begin{matrix} 1 & 1 & 1 \cr m^2_1 + m^2_3 & m^2_2 + m^2_3 & m^2_1 + m^2_2 \cr m^2_1 m^2_3 & m^2_2 m^2_3 & m^2_1 m^2_2\end{matrix}\right) \cdot \left(\begin{matrix} h^2_{13} m^{}_1 m^{}_3 \Delta^{}_{13} & 0 & 0 \cr 0 & h^2_{23} m^{}_2 m^{}_3 \Delta^{}_{23} & 0 \cr 0 & 0 & h^2_{12} m^{}_1 m^{}_2 \Delta^{}_{12}\end{matrix}\right) \; ,
%     (31)
\end{eqnarray}
where $\Delta^{}_{ij} = m^2_i - m^2_j$ for $i,j = 1, 2, 3$ have been defined. Then it is easy to confirm the result for the determinant of ${\cal A}$ in Eq.~(\ref{eq:det}). In contrast, the expressions of $g^{}_i$ (for $i = 1, 2, \cdots, 6$) relevant for the WB invariant ${\cal I}^{}_3$ are rather lengthy. However, we list them below for completeness:
\begin{eqnarray}
g^{}_1 &=& +6{\rm i}m^{}_1 m^{}_3 \left\{\left[m^4_2 e (be - ac) + m^2_1 m^2_3 b (df - b^2) \right] \cdot \left[b(c^2 - a^2) + ac (d - f)\right] \right.\nonumber \\
&~& + m^2_1 m^2_2 \left\{ ab(ab - cd)(b^2 - a^2) + d \left[ b (c^2 - a^2) + ac (d - e)\right](ac - be) \right. \nonumber \\
&~& \left. + a b (e - f) \left[ab (d - e) + c (a^2 - d^2)\right] \right\} \nonumber \\
&~& + m^2_2 m^2_3 \left\{ bc (bc - af)(c^2 - b^2) + f \left[ b (c^2 - a^2) + ac (e - f)\right](ac - be) \right. \nonumber \\
&~& \left.\left. + b c (e - d) \left[ bc (e - f) + a (f^2 - c^2)\right] \right\} \right\} \; ,
\end{eqnarray}
\begin{eqnarray}
g^{}_2 &=& -6{\rm i}m^{}_2 m^{}_3 \left\{\left[m^4_1 d (cd - ab) + m^2_2 m^2_3 c (ef - c^2) \right] \cdot \left[c(a^2 - b^2) + ab (f - e)\right]\right. \nonumber \\
&~& + m^2_1 m^2_3 \left\{ bc(bc - af)(c^2 - b^2) + f \left[ c (a^2 - b^2) + ab (f - d)\right](ab - cd) \right. \nonumber \\
&~& \left. + b c (d - e) \left[bc (f - d) + a (b^2 - f^2)\right] \right\} \nonumber \\
&~& + m^2_1 m^2_2 \left\{ ac (ac - be)(a^2 - c^2) + e \left[ c (a^2 - b^2) + ab (d - e)\right](ab - cd) \right. \nonumber \\
&~& \left.\left. + a c (d - f) \left[ ac (d - e) + b (e^2 - a^2)\right] \right\} \right\} \; ,
\end{eqnarray}
\begin{eqnarray}
g^{}_3 &=& -6{\rm i}m^{}_1 m^{}_2 \left\{ \left[m^4_3 f (af - bc) + m^2_1 m^2_2 a (de - a^2) \right] \cdot \left[a(b^2 - c^2) + bc (e - d)\right]\right. \nonumber \\
&~& + m^2_2 m^2_3 \left\{ ac(ac - be)(a^2 - c^2) + e \left[ a (b^2 - c^2) + bc (e - f)\right](bc - af) \right. \nonumber \\
&~& \left. + a c (f - d) \left[ac (e - f) + b (c^2 - e^2)\right] \right\} \nonumber \\
&~& + m^2_1 m^2_3 \left\{ ab (ab - cd)(b^2 - a^2) + d \left[ a (b^2 - c^2) + bc (f - d)\right](bc - af) \right. \nonumber \\
&~& \left.\left. + a b (f - e) \left[ ab (f - d) + c (d^2 - b^2)\right] \right\} \right\} \; ,
\end{eqnarray}
and
\begin{eqnarray}
g^{}_4 &=& +6{\rm i}m^{}_1 m^{}_2 m^2_3 \left[b(ab - cd)m^2_1 - c(ac - be) m^2_2\right] \cdot \left[a(c^2 - b^2) + bc(d - e)\right] \; , \nonumber \\
g^{}_5 &=& +6{\rm i}m^{}_1 m^2_2 m^{}_3 \left[a(ab - cd)m^2_1 - c(bc - af) m^2_3\right] \cdot \left[b(c^2 - a^2) + ac(d - f)\right] \; , \nonumber \\
g^{}_6 &=& +6{\rm i}m^2_1 m^{}_2 m^{}_3 \left[a(ac - be)m^2_2 - b(bc - af) m^2_3\right] \cdot \left[c(b^2 - a^2) + ab(e - f)\right] \; ,
\end{eqnarray}
where $a \equiv h^{}_{12}$, $b \equiv h^{}_{13}$, $c \equiv h^{}_{23}$, $d \equiv h^{}_{11}$, $e \equiv h^{}_{22}$ and $f \equiv h^{}_{33}$ have been defined to just simplify the notations. It should be stressed that although the expressions of $g^{}_i$ (for $i = 1, 2, \cdots, 6$) are very complicated, they are indeed indpendent of the Majorana CP phases $\rho$ and $\sigma$.

\end{appendix}

\end{document}